\documentclass[onecolumn,draftclsnofoot]{IEEEtran}

\usepackage{times,amsmath,epsfig,latexsym,amssymb,psfrag}
\usepackage{booktabs}
\usepackage{citesort}
\usepackage{paralist}
\usepackage[ruled,vlined]{algorithm2e}
\usepackage{float}
\restylefloat{table}

\usepackage{float}
\floatstyle{plaintop}
\restylefloat{table}

  \usepackage{setspace}
\newcommand{\qed}{\nobreak \ifvmode \relax \else
\ifdim\lastskip<1.5em \hskip-\lastskip
\hskip1.5em plus0em minus0.5em \fi \nobreak
\vrule height0.75em width0.5em depth0.25em\fi}

\begin{document}

\title{Energy Efficient Scheduling and Grouping for Machine-Type Communications over Cellular Networks}
\author{ Amin Azari\\
KTH Royal Institute of Technology\\
Email: aazari@kth.se}
\maketitle


\begin{abstract}
In this paper, energy-efficient scheduling for grouped  machine-type devices deployed in cellular networks is investigated. We introduce a scheduling-based  cooperation incentive scheme which enables machine nodes to organize  themselves locally, create machine groups, and communicate through group representatives to the base station. This scheme benefits from a novel scheduler design which takes into account the cooperation level of each node, reimburses the extra energy consumptions of group representatives, and maximizes the network lifetime. As reusing cellular uplink resources for communications inside the groups degrades the Quality of Service (QoS) of the primary  users, analytical results are provided which present a tradeoff between  maximum allowable number of simultaneously  active machine groups in a given cell and QoS of the primary users.
Furthermore, we extend our derived solutions for the existing cellular networks, propose a cooperation-incentive LTE scheduler, and present our simulation results in the context of LTE. The simulation results show that the proposed solutions  significantly prolong the network lifetime. Also, it is shown that under certain circumstances, reusing uplink resource by machine devices can degrade the outage performance of the primary users significantly, and hence, coexistence management of machine devices and cellular users is of paramount importance for next generations of cellular networks in order to enable group-based machine-type communications while guaranteeing  QoS for the primary users.

\end{abstract}

\IEEEpeerreviewmaketitle

 \section{Introduction}
Information and Communications Technology (ICT) can play an important role in smart cities in order to improve the services that support
urban dwelling like security, healthcare, public
transportation, remote working and education, entertainment, and communications \cite{fair,green}. For example, due to ever increasing portion of old-aged people in the societies, ICT can provide a wide-range of health-care applications in order to monitor the status of citizens right from their homes by smart sensors \cite{hel}.  The number of smart devices is expected to be nearly 50
billion by 2020, based on the estimation from Ericsson \cite{rico}, and these devices are expected to be able to communicate autonomously. 
Machine-to-Machine (M2M) communications means the inter-communications of machine devices without human intervention, and aims at enabling ubiquitous connectivity among uniquely identifiable smart physical objects that are capable of sensing or acting on their environment \cite{m2m1}.
An important enabler for M2M communications over cellular networks is Machine-Type Communications (MTC) which is expected to play a critical role in the market since cellular networks are penetrated deeply into almost all locations, provide easy-to-use cost-efficient communications anywhere by ubiquitous  coverage and roaming capability \cite{itu}\nocite{w_sony}-\cite{w_sam}. M2M networks are generally characterized by the massive number of concurrent active devices, low-payload data transmission, and vastly diverse Quality-of-Service (QoS) requirements \cite{ghav}.
The continuing growth in demand from machine-type communications \cite{cisco}, the fact that most smart devices are battery driven and long battery-lifetime is crucial for them, and the inefficiency in current cellular infrastructure for providing energy-efficient small data communications \cite{llte} have triggered many research projects to see how current cellular standards must  be revisited in order to provide large-scale yet efficient machine-type communications \cite{8l}-\nocite{9l}\nocite{how}\nocite{m2monlte}\nocite{how1}\cite{sched}. 

In this paper, energy efficient scheduling for grouped machine-type devices over cellular networks is investigated. Aiming at maximizing the network lifetime, we present a cooperation incentive scheduler which  reimburses the extra energy consumptions for the helper nodes. Also, we extend the derived  solutions for existing cellular networks. Finally, we analyze the impact of underlying intra-group communications on the uplink transmission of primary users.

The main contributions of this paper include:
\begin{itemize}

\item
Present a lifetime-aware cooperative machine-type communications framework for future cellular networks with dense MTC deployment.
\item
Present a cooperation-aware scheduler which reimburses the extra energy consumptions of the helper nodes and maximizes the network lifetime. Investigate the application of the proposed scheduler for existing LTE networks.
\item
Present a distributed grouping scheme for machine-type  devices deployed in cellular networks. 
\item
Explore analytically the tradeoffs between maximum allowable number of  overlay machine groups in a cell and the interference level at the primary user.
\end{itemize}

The remainder of this article is organized as follows. Section II presents the related works. The system model is introduced in section III. In section IV, the cooperation incentive scheduler  is presented. In section V, distributed grouping  is presented. In section VI, a cooperation-incentive MTC scheduler for LTE networks is investigated. The impact of intra-group M2M communications on QoS of primary users is investigated in section VII.    Simulation results are given in section VIII. Concluding remarks are presented in  section IX.


\section{Related Works and Motivation}
\subsection{Cellular-access for machine-type subscribers }
The continuing growth in demand from machine-type subscribers for small data  communications  poses significant challenges to the existing cellular networks \cite{ghav}.  Random Access Channel (RACH)  in the Long-Term Evolution (LTE) networks  is the typical way for machine nodes to connect to 
the BS \cite{andres}. Regarding the limited number of available preambles per time per cell, collisions and energy wastages are likely to happen  when a large number of machine-type devices try to connect to the BS \cite{sched}. Several solutions, including Access Class Barring (ACB) \cite{acb1,acb2}, prioritized random access \cite{pri,pri1}, and connectionless MTC \cite{con1,con2} are studied in literature to reduce congestion in an overload condition. 
 In ACB, each machine node in class $i$ which  has data to transmit,  decides  to content for channel access with probability $p_i$ and  to defer the transmission with $1-p_i$.  In prioritized random access schemes \cite{pri,pri1}, the available preambles are divided among different classes of users, and hence, each class of users has only access to a limited set of preambles. To further reduce the probability of congestion in the RACH of LTE, machine nodes with limited data packet size can send data directly to the BS  without connection establishment \cite{con1,con2}. 

The surge in the number of connected devices \cite{r10}, not only affects the Radio 
Access Network (RAN) of LTE but also severe signaling overload is expected to happen at the Evolved Packet Core (EPC) of LTE \cite{r11}. In \cite{sca}, the scalability of MTC and Internet of Things (IoT) on existing LTE networks is investigated and it is shown that some MTC traffic categories, e.g. asset tracking, can be substantially challenging.   Then, efficient scheduler design with reduced signaling overhead is of paramount importance for future M2M-enabled cellular networks \cite{sched}. 



\subsection{Towards energy efficient MTC over cellular networks }
To curb undesirable energy wastage and extend the battery lifetime of machine devices, a variety of energy-conserving measures have been developed, including Discontinues Reception (DRX), group-based MTC, and 
energy-efficient MTC scheduling.

\subsubsection{ Discontinues reception}
DRX has been specified in LTE for energy saving by allowing User Equipment (UE) to switch off its receiver and go to the sleep mode. During the sleep period, UE cannot receive packets  from the base station, and hence, DRX can  introduce delays in data reception. To maximize the energy saving gain and keep the aforementioned delay bounded, the DRX parameters are to be designed carefully \cite{dtx}.
DRX design for machine-type communications is investigated in \cite{sach,dtx1}, and it is shown that such operation can significantly prolong the battery lifetimes of machine nodes deployed in LTE networks. 

\subsubsection{Group-based MTC}
Cooperation among machine devices and a group-based operation of them seem to be promising approaches for offloading the BS \cite{ghav}. In addition to the BS offloading, our previous work in \cite{gsip} showed that network-assisted grouping can significantly prolong  the M2M network lifetime.  

\subsubsection*{Group-based MTC through gateways}
Integrating    MTC  Gateways  in    cellular network  architecture  enables us to handle a massive  number of  concurrent access requests using capillary networking, and  extend  the  coverage  in  remote areas \cite{ghav,sched}. 
Group-based MTC has been studied in \cite{sachs}, and the authors claim that the capillary networks will become a key enabler of the future networked society. Addressing the massive machine access problem for existing cellular networks with the help of M2M gateways and capillary networks is  investigated in \cite{gw1}. 

 
\subsubsection*{Group-based MTC through UEs}
D2D communications over cellular networks motivates the idea to  relay the MTC traffic originated from the machine nodes through the D2D links \cite{grlaya}. 
MTC traffic relaying through D2D links underlying downlink transmission of primary cellular users is investigated in \cite{d2ddo}. Aggregation of locally generated MTC traffic through D2D links is investigated in \cite{d2dagg}, where a tradeoff between latency and the transmit power, which is needed to deliver the aggregate traffic, is presented. A multi-hop routing scheme for MTC traffic consisting of opportunistic D2D links is presented in \cite{d2dmh}.  
  \subsubsection*{Group-based MTC through M2M links}
Without a fixed a priory installed M2M gateway or a cellular UE which is eager/secure to relay the MTC traffic, each machine node could temporary act as a GR \cite{grlaya}. In this case, the selection of the GR should be based on the remaining energy level of machine nodes and the channel state with the BS. Energy-efficient clustering and medium access control (MAC) design for a massive number of machine devices which connect to the BS in the asynchronous mode, e.g. RACH of LTE, is considered in \cite{gsip,vtc,jsac}, where feasible regions for clustering, the optimal cluster-size, cluster-head selection, and  energy-efficient communications protocol design are investigated. 
\subsubsection{Energy efficient MTC scheduling}\label{tcms}
 The energy-efficient uplink scheduling in LTE networks with coexistence of machine- and human-oriented traffic is investigated in \cite{coex}. Power-efficient Resource Allocation (RA)  for MTC is developed in \cite{poshti,ra}. 
To facilitate network design, the 3rd Generation Partnership Project (3GPP) and IEEE have defined specific service requirements  for M2M communications where one of the most important ones is the time-controlled feature \cite{sysreq,m2meee}. This feature is intended for use with M2M applications that can tolerate to send data only during defined time intervals. 
Based on this feature, a reduced complexity approach for scheduling a massive number of machine devices  is introduced in \cite{mas}. This scheme organizes machine devices with similar QoS requirements into classes
where each class is associated with a prescribed QoS profile, including priority level, packet delay and  jitter budget, and dropped packet rate requirement.
Then, fixed access grant time intervals are allocated to each
class based on the traffic rate and the priority of each class. This time-controlled scheduling framework for semi-constant rate machine devices is widely adopted in literature \cite{sched}, \cite{ana}-\nocite{grp1}\nocite{grp3}\nocite{clasque}\nocite{tree}\cite{reli}.  In \cite{wcnc,vtc}, we have adopted this time-controlled scheduling framework, presented an accurate energy consumption model for MTC communications, and presented lifetime-aware scheduling solutions for SC-FDMA-based networks.




\subsection{Motivation}
There are many M2M applications that require very high energy efficiency to ensure long battery lifetime. 
While the proposed scheduling schemes in \cite{poshti,ra,wcnc,vtc} aim at enabling energy-efficient machine-type communications, network congestion including radio network congestion and signaling network congestion as defined in \cite{m2mm2} is likely to happen in serving a massive number of machine devices. Relaying MTC traffic over fixed gateways and D2D links are promising approaches to prevent such kind of congestions  in cellular networks \cite{mc1,d2dmh}. However, when a fixed a priory installed M2M gateway is unavailable, or a cellular UE which is eager and secure to cooperate with machine nodes is unavailable, a machine node can relay its neighbors' packets to the BS. Here, the  energy costs for the helper nodes is of paramount importance because they are also battery-limited machine-type devices and wish to maximize their individual lifetimes. These problems motivate us to propose a novel scheduler for  grouped MTC scheduling over cellular networks.


\section {System Model}\label{sys}
 Consider a single cell with one base station in the center and a massive number of machine nodes which are uniformly deployed in the cell. The machine nodes are battery-limited, and hence, long battery-lifetime is crucial for them.   
 For node $i$, denote the remaining energy at time $t_0$ by $E_i(t_0)$. Define the power consumption in transmitting mode for node $i$ as $\alpha P_i+P_c$ where $P_c$ is the circuit power consumed by electronic circuits in transmission mode, $\alpha$ is the inverse of power amplifier efficiency, and $P_i$ is the transmit power for reliable data transmission. Also, denote by  by $E_s^i$ the average static energy consumption in each duty cycle for data gathering, synchronization, admission control, and etc. Then, the expected lifetime for node $i$ at time $t_0$ can be expressed as the product of duty cycle and the ratio between remaining energy at time $t_0$ and the required energy consumption in each duty cycle, as follows \cite{wcnc}:
\begin{equation}\label{eq.L}
L_i(t_0)=\frac{E_i(t_0)}{E_s^i+\tau_i(P_c+\alpha P_i)}T_i.
\end{equation}
In this expression, $T_i$ is the expected length of each duty cycle, $\tau_i$ is the data transmission time. Given the individual lifetime of machine nodes, one may define the network lifetime  as a function of individual lifetimes in different ways \cite{wcnc}. In this paper we consider {\it First Energy Drain} (FED) network lifetime which refers to the time at which first node drains out of energy, and is applicable when missing even one node deteriorates the performance or coverage.

%
%
%
%
%

\section{Scheduling As a Cooperation Incentive Strategy}\label{cop}
Consider the system model in section \ref{sys} where a set of  machine devices $\pounds$ with cardinality $\mathcal L$ must be scheduled at once.   The results for other multiple-access schemes can be similarly derived.  Denote the total number of available resource elements as $c_t$, the length of each resource element as $\tau_r$, and the allocated fraction of time  for data transmission of node $i$ as $\tau_i=c_i\tau_r$.  Denote the pathloss between node $i$ and the BS as $g_i$, and the noise power spectral density (PSD) at the receiver as $N_0$. Then, using the Shannon capacity formula the transmit power of node $i$  is derived as a function of $c_i$ as follows:
\begin{equation}\label{powe}
P_i={G_i}(2^{\frac{D_i}{c_i\tau_r w}}-1), \forall i\in \pounds,
\end{equation}
where  $G_i=g_i \Gamma N_0w$, $w$ is the bandwidth, and $\Gamma$ is the signal to noise ratio (SNR) gap between channel capacity and a practical coding and modulation scheme. As transmit power of each device is bounded by $P_{\max}$, the lower bound on $c_i$ is found as:
\begin{equation}\label{cmin}
c_i^{min}=\left\lceil {\frac{D_i}{\tau_r w\log_2(1+ { P_{\max}}/{G_i})}} \right\rceil, \forall i\in\pounds.
\end{equation}
 Instead of scheduling all machine nodes to connect directly to the core network, the BS may prefer to promote group-based communications, especially in the case of dense machine deployment. Grouping has been proved to be effective to reduce collisions and increase network-level energy efficiency in M2M-enabled cellular networks \cite{istiak,sched,sachs}. With grouping, a selected node/gateway is responsible for relaying group member's packets to the base station. As a result, the energy consumption of the helper nodes will be much higher than the other nodes. Here, we present a cooperation  incentive strategy which utilizes uplink scheduling as a rewarding mechanism to reimburse the extra energy consumptions in the helper nodes. In our scheme, BS  broadcasts a promotion message in the cell, which includes  a rewarding parameter $\beta$. From  this parameter, each node  can determine its optimal strategy, i.e. to make a group or attach to another group, and the optimal group size. After group forming, to be discussed in the section \ref{dgf}, the initial set of nodes reduces to $\pounds_r$ with cardinality ${\mathcal L}_r$.  Designing  intra-group communications protocols is out of the scope of this paper and is left as a research direction in which this work can be extended. For preliminary results, the interested reader may refer to  \cite{gsip,jsac}, where a hybrid communications protocol for intra-group communications is investigated. In the following, we focus on formulating a resource allocation problem which benefits from dealing with a lower number of devices, considers the cooperation level of the helper nodes, and maximizes the network lifetime. Then, the joint scheduling and power control optimization problem that maximizes the network lifetime  is written as follows:

\begin{align}
 \max_{\beta,c_i, P_i} &\hspace{2mm} \min_{i\in \pounds_r}\hspace{1mm} L_i(t_0) \label{op150}\\
\text{subject to: }&\text{C}.\ref{op150}\text{.1: }\sum\nolimits_{i\in\pounds_r}c_i\le c_t,\nonumber\\
 &\text{C}.\ref{op150}\text{.2: } c_i^{m}\le c_i, \hspace{1mm}\quad\forall i\in {\bf{\pounds_r}},\nonumber\\
&\text{C}.\ref{op150}\text{.3: } c_i^{r}\hspace{0.5 mm} \text{resource elements are guaranteed for node}\hspace{0.5 mm} i, \nonumber\\
&\text{C}.\ref{op150}\text{.4: }\sum\nolimits_{i\in {\bf{\pounds_r}}} ( \beta n_i+1)c_i^{min}\le c_t \nonumber,
\end{align}
where $c_i^{r}=( \beta n_i+1) c_i^{min} $, $c_i^{m}$ is found from \eqref{cmin} as:
\begin{align}
c_i^{m}=\left\lceil {\frac{(n_i+1){\hat D}}{\tau_r w\log_2(1+ { P_{\max}}/{G_i})}} \right\rceil,
\end{align}
$n_i$ is the number of clients for node $i$, $E_{h}$ the energy consumption in listening mode for data collection from a neighbor node, ${\hat D}$ the average packet size over the set of connected machine nodes, and $L_i(t_0)$ is found by rewriting  the lifetime expression in \eqref{eq.L}  as follows:
\begin{equation}\label{lif2}
L_i(t_0)=\frac{E_i(t_0)}{E_s^i+\tau_i(P_c+\alpha P_i)+n_iE_h}T_i.
\end{equation} 
One must note that beside network lifetime, other performance measures, to be discussed in section \ref{dgf}, also contribute in determining the optimal $\beta$. Then, in the reminder of this paper we assume that $\beta$ is known a priory at the BS. The optimal choice of $\beta$ is left as a research direction in which this work can be extended. Let us first solve the optimization problem in \eqref{op150} by assuming that $\beta=0$, i.e. when scheduler is not aware of the cooperation level of each node. When $\beta=0$, one can rewrite the optimization problem in \eqref{op150} as:
\begin{align}
\max_{c_i, P_i}& \hspace{2mm} \min_{i\in \pounds_r} L_i(t_0)\label{op0}\\
  \text{subject to: }&\text{C}.\ref{op0}\text{.1: }\nonumber\sum\nolimits_{i\in \pounds_r}c_i \le c_t,\nonumber\\
&\text{C}.\ref{op0}\text{.2: } c_i^{m}\le c_i   \quad\forall i\in\pounds_r.\nonumber
\end{align}
Using linear relaxation, the integer scheduling problem in \eqref{op0} transforms into a related linear optimization problem, as follows:

\begin{align}
\max_{\tau_i}& \hspace{2mm} \min_{i\in \pounds_r} L_i(t_0)\label{op1}\\
  \text{subject to: }&\text{C}.\ref{op1}\text{.1: }\nonumber\sum\nolimits_{i\in \pounds_r}\tau_i \le c_t \tau_r,\nonumber\\
&\text{C}.\ref{op1}\text{.2: }\tau_i^{m}\buildrel \Delta \over =\frac{(n_i+1){\hat D}}{w\log_2(1+ { P_{\max}}/{G_i})} \le \tau_i  \quad\forall i\in\pounds_r.\nonumber
\end{align}
By inserting \eqref{powe} in \eqref{lif2}, one  sees that the lifetime expression $L_i(t_0)$ in \eqref{op1} and its inverse $L_i^{-1}(t_0)$ are strictly quasiconcave and convex functions of $\tau_i$ respectively. Then, one can transform the problem in \eqref{op1} to a related convex optimizations problem by defining $z$ as an auxiliary variable where $z=\max_{i\in{\bf{\pounds_r}}}\hspace{2mm}L_i^{-1}(t_0)$,
 and rewriting \eqref{op1} as follows:
\begin{align}
\min_{\tau_i}& \hspace{2mm} z \label{op2}\\
\text{subject to:}&\hspace{1.5 mm} \text{C}.\ref{op1}\text{.1}, \text{C}.\ref{op1}\text{.2}, \text{and}\nonumber\\
 &L_i^{-1}(t_0)\le z, \quad \forall i\in {\bf{\pounds_r}}\nonumber.
\end{align}
As $L_i^{-1}(t_0)$ is a convex function of $\tau_i$, and $z$ is a convex 
function\footnote{Because the point-wise maximum operation preserves convexity \cite{boyd_con}.} of $\tau_i$, the joint scheduling and power control problem in \eqref{op2} is a convex optimization problem. Using the method of Lagrange multipliers \cite{boyd_con}, the optimal transmission time for node $i$ is found as follows:
\begin{align}
\tau_i^s&= \max\bigg\{\frac{\ln(2){(n_i+1){\hat D}}}{w\times \text{lambertw}(\frac{-1}{\mathrm e}+\frac{{P_c  \lambda_i}+{ T_i\mu E_i(t_0) }}{{G_i\lambda_i}{ \mathrm e}{\alpha} })+w}  , \tau_i^{m}\bigg\}, \label{tis}
\end{align}
where  $\mu$ and $\lambda_i$:s are Lagrange multipliers, $\tau_i^{m}$ has been introduced in \eqref{op1}, the lambertw function\footnote{Also called the omega function or product logarithm.} is the inverse  of the function $f(x) = xe^x$ \cite{lam}, and $\mathrm e$ is the Euler's number. Also, the Lagrange multipliers are found due to the following Karush Kuhn Tucker\footnote{First order necessary conditions for a solution in nonlinear programming to be optimal.} (K.K.T.) conditions \cite{boyd_con}:
\begin{align}
\mu\ge 0; \quad \lambda_i&\ge 0 \quad \forall i\in \pounds_r; \nonumber\\
 \mu(\sum\nolimits_{i\in \pounds_r}&\tau_i-c_t \tau_r)=0;\nonumber\\
  \lambda_i({L_i^{-1}(t_0)}-&z)= 0 \quad \forall i\in \pounds_r.\label{kkt}
\end{align}
To find the number of allocated resource elements to node $i$, one can divide $\tau_i^s$ by $\tau_r$. The  $\tau_i^s$ expression in \eqref{tis} is a fractional solution to the relaxed problem and the $\frac{\tau_i^s}{\tau_r}$ expression is  not necessarily integer. Then, one can use randomized rounding to find the number of assigned resource elements to each node, i.e. $c_i^s$, and also satisfy the  integrality constraint \cite{round}.   Finally, the uplink transmit power of node $i$ is found from \eqref{powe} where $c_i=c_i^s$.

Now, let come back to the original problem in \eqref{op150} where $\beta > 0$. One can transform this problem  to a convex optimizations problem by defining $z=\max_{i\in{\bf{\pounds_r}}}\hspace{2mm}{L_i^{-1}(t_0)}$ as an auxiliary variable
 and rewriting \eqref{op150} as follows:
\begin{align}
\min& \hspace{2mm} z \label{op15}\\
\text{subject to: }&\text{C}.\ref{op15}\text{.1: }\sum\nolimits_{i\in\pounds_r}\tau_i\le c_t \tau_r,\nonumber\\
 &\text{C}.\ref{op15}\text{.2: } \tau_i^{m}\le \tau_i, \hspace{1mm}\quad\forall i\in {\bf{\pounds_r}},\nonumber\\
&\text{C}.\ref{op15}\text{.3: }{L_i^{-1}(t_0)}\le z, \quad \forall i\in {\bf{\pounds_r}}\nonumber,\\
&\text{C}.\ref{op15}\text{.4: } \tau_i^{r}\hspace{0.5 mm} \text{seconds are guaranteed for node}\hspace{0.5 mm} i \nonumber,
\end{align}
where  
\begin{align}
  \tau_i^{r}&=( \beta n_i+1) \tau_i^{min},\nonumber\\
  \tau_i^{min}&= {\frac{{ D_i}}{ w\log_2(1+ { P_{\max}}/{G_i})}} ,\nonumber\\  
 P_i&=(2^{\frac{(n_i+1)\hat{D}}{\tau_i w}}-1){G_i}\nonumber.
\end{align}
To unify the  second and fourth constraints, we can formulate a subproblem as:
\begin{align}
\tau_i^*= \arg&\max_{\tau_i} \hspace{1mm}L_i(t_0)\nonumber \\
=\arg &\min_{\tau_i} \hspace{1mm}\tau_i(P_c+\alpha (2^{\frac{ (n_i+1){\hat D}}{w\tau_i}}-1){G_i}\label{eqn}\\
\text{subject to:}&\hspace{2mm} \tau_i^{m}\le\tau_i\le\tau_i^{r}. \nonumber
\end{align}
The objective function in \eqref{eqn} is a strictly convex function of $\tau_i$ and chooses its minimum value at 
\begin{equation}\label{mv}\tau_i^{x}=\min\{\max\{\tau_i^m ,\frac{\ln(2) { (n_i+1){\hat D/w}}  }{ \text{lambertw}( \frac{P_c  -\alpha G_i }{\mathrm e G_i \alpha {}}) + 1}\},\tau_i^r\}.\end{equation}
Then, one can rewrite the optimization problem in \eqref{op15} as:
\begin{align}
\min& \hspace{2mm} z \label{op160}\\
\text{subject to: }&\text{C}.\ref{op15}\text{.1}, \text{C}.\ref{op15}\text{.3}, \text{and}\nonumber\\ 
 &\tau_i^{x}\le \tau_i, \hspace{1mm}\quad\forall i\in {{\pounds_r}}.\nonumber
\end{align}
One sees that this problem is similar to the optimization problem in \eqref{op2}. Then, one can pursue the same approach to prove that  \eqref{op160} is a convex optimization problem. Using\eqref{tis}, one can derive the optimal solutions for \eqref{op160} as follows:
\begin{align}
\tau_i^*&= \max\bigg\{\frac{\ln(2){(n_i+1){\hat D}/w}}{ \text{lambertw}(\frac{-1}{\mathrm e}+\frac{{P_c  \lambda_i}+{ T_i\mu E_i(t_0) }}{{G_i\lambda_i}{ \mathrm e}{\alpha} })+1}  , \tau_i^x \bigg\}, \label{ti*}
\end{align}
where the Lagrange multipliers are found due to the K.K.T. conditions in \eqref{kkt}. To find the number of allocated resource elements to node $i$, one can divide $\tau_i^*$ by $\tau_r$, and use randomized rounding to find the number of assigned resource elements to each node, i.e. $c_i^*$.   Finally, the uplink transmit power of node $i$ is found from \eqref{powe} where $c_i=c_i^*$.

\section{Distributed rewarding-based grouping}\label{dgf}
To enable group-based MTC, a distributed  grouping scheme is needed to help machine nodes organize themselves locally and form machine groups. Regarding our proposed scheme in section \ref{cop}, BS broadcasts the rewarding parameter $\beta$, and each node independently decides to broadcast itself as a GR or attach to the group of its nearest GR. In other words, we assumed that given the incentive parameter $\beta$, each node is able to derive its optimal number of clients $n_i^*$. In this section, we settle this issue by formulating the group forming as an optimization problem which can be solved locally at each node. 
If node $i$ broadcasts itself as a GR, $\tau_i^{r}=(\beta n_i+1)\tau_i^{min}$ seconds will be guaranteed for it. Then, one can formulate the grouping  problem as a lifetime-maximization problem as follows:
\begin{align}
\max_{\tau_i, n_i} & \hspace{2mm} {L_i(t_0)} \label{op16}\\
\text{subject to:  }
  & \text{C}.\ref{op16}\text{.1:}\hspace{1mm}  \tau_i^m\le \tau_i\le \tau_i^{r},\nonumber\\
& \text{C}.\ref{op16}\text{.2:}\hspace{1mm} n_i\in\{0,1,\cdots, n_{max}\},\nonumber
\end{align}
where $n_{max}$ is limited due  the practical limits on the maximum number of sustained clients for a machine node. By manipulating the objective function, the optimization problem in \eqref{op16} reduces to: 
\begin{align}
\min & \hspace{2mm}{\tau_i P_c+\tau_i \alpha (2^{\frac{(n_i+1){\hat D}}{\tau_i w}}-1) {G_i }+n_i E_h} \label{op17}\\
\text{s.t.:}& \hspace{2mm}\text{C}.\ref{op16}\text{.1}, \text{C}.\ref{op16}\text{.2}.\nonumber
\end{align}
One sees that the objective function in \eqref{op17} is a strictly convex function of $n_i$ and $\tau_i$, but it is not jointly convex.  If we fix $n_i$, the optimal transmission time for a given $n_i$, i.e. $\tau_i^*(n_i)$, is found from the \eqref{eqn}, as follows:
$$\tau_i^*(n_i)=\min\{\max\{\tau_i^m ,\frac{\ln(2) { (n_i+1){ D/w}}  }{ \text{lambertw}( \frac{P_c  -\alpha G_i }{\mathrm e G_i \alpha {}}) + 1}\}.
$$
Now, the optimization problem in \eqref{op17} reduces  to the following optimization problem:
\begin{equation}\label{op18}
n_i^*=\arg \max_{n_i\in\{0,\cdots,n_{max}\}} \hspace{1mm} L_i(t_0)\big|_{\tau_i=\tau_i^{*}(n_i)}. 
\end{equation}
If the incentive parameter $\beta$ is known, each node can solve this optimization problem independently, and find its optimal number of clients. For the $i$th node, if $n_i^*> 0$ then it can broadcast itself as a GR. If the number of received requests from neighbor nodes which prefer to connect  to node $i$, i.e. $n_i^{r}$, is less than $n_i^*$, then this node can derive its optimal number of clients by substituting $n_{max}$ with $n_i^{r}$, and once again solving the optimization problem in \eqref{op18}. 

 One must note that beside network lifetime, other performance measures like level of interference to the primary user may also contribute in determining the optimal $\beta$. For example, in section \ref{sim} we will show how the choice of $\beta$ impacts the average number of clients per each GR, and hence, the average number of groups in the cell. Also, in section \ref{int} we will present how the number of active groups in the cell, which are reusing the uplink resources, can impact the QoS for a primary user.  The BS can use learning-based approaches \cite{lear}, and by monitoring the level of interference at the primary users, determine the optimal choice of $\beta$. The optimal choice of $\beta$ is left as a research direction in which this work can be extended.

\section{ Cooperation Incentive Scheduler for LTE networks}\label{ltesim}
\begin{figure}[!t]
\centering
\includegraphics[trim={1cm .1cm 0cm .2cm},clip,width=5in]{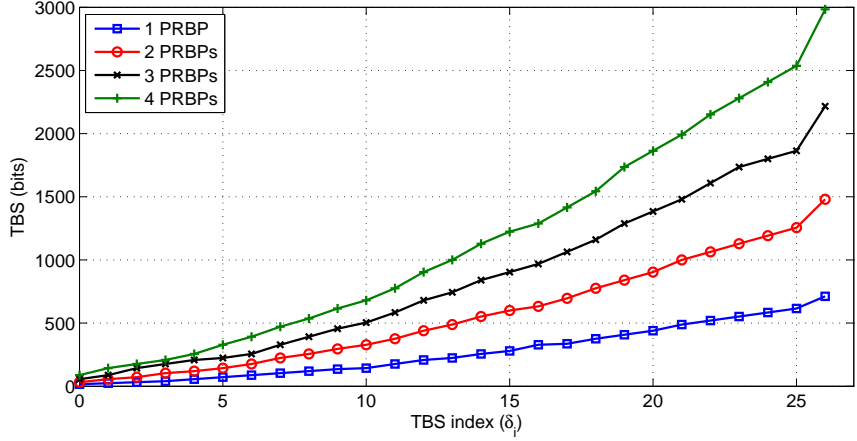}
\caption{ Transport block size as a function of $\delta_i$ and number of PRBPs} \label{tbs}
\end{figure}

In order to investigate possible benefits of our proposed solutions in practice, here we extend our derived solutions for the existing LTE networks.  LTE is a standard for high-speed wireless data communications of mobile phones and data terminals. This  standard is developed by the 3GPP, and supports deployment on different frequency bandwidths, including: 1.4MHz, 3MHz, 5MHz, 10MHz, 15MHz, and 20MHz.  Let us consider the air interface of the 3GPP LTE Release 10 \cite{3gpp}, where radio resources for uplink transmission are distributed in both time and frequency domains. Data transmission in time domain is structured in frames where each frame consists of 10 subframes each with 1 ms length. In the frequency domain, the available bandwidth is divided into a number of subcarriers each with 15 KHz bandwidth.
A Physical Resource Block Pair (PRBP) is the minimum allocatable resource element in an LTE frame which consists of 12 subcarriers  spanning over one Transmission Time Interval (TTI) \cite{3gpp}. Each TTI consists of two slots and includes 12  OFDM symbols if long cyclic prefix is utilized. 
Denote the number of assigned PRBPs to node $i$ by $c_i$,  the estimated downlink pathloss by $\gamma_i$, the compensation factor by $\beta_i$, the number of symbols in a PRBP by $N_s$, the noise power in each resource block by $p_n$, and the required SNR level at the receiver by $\gamma_0$. Then, the uplink transmit power of each node is determined using the LTE open-loop uplink power-control mechanism in \cite{3gpp1}, as follows:
 \begin{align}
P_i = c_i\big(\beta_i(\gamma_0+p_n)+(1-\beta_i P_{max})\big) \beta_i \gamma_i(2^{\frac{1.25\text{TBS ($c_i$, $\delta_i$)}}{12 c_iN_s}}-1),\label{pil}
\end{align}
where the Transport Block Size (TBS) can be found in Table 7.1.7.2.1-1 of \cite{3gpp1} as a function of $c_i$ and TBS index. For a quick reference, the first 4 columns of this table are depicted in Fig. \ref{tbs}. The TBS index, $\delta_i \in\{0,\cdots,26\}$, is a function of modulation and coding scheme as  in Table 8.6.1-1 of \cite{3gpp1}.  
Regarding the aforementioned scheduling framework in LTE, one can use the derivations in section \ref{cop} and rewrite the expected lifetime for node $i$ as follows:
\begin{align}
L_i(t_0)=\frac{E_i(t_0)T_i}{E_s^i+n_iE_h+TTI(P_c+\alpha P_i)}.\label{lil}
\end{align}
Aiming at FED network lifetime maximization, uplink scheduling for machine-type communications over LTE networks is investigated in \cite{vtc}. Inspired by  the  derived solutions in section \ref{cop}, we modify the proposed algorithm in \cite{vtc}, and present a cooperation incentive scheduler  for MTC over LTE networks, as in Algorithm \ref{alg1}.  In this algorithm, the cooperation incentive parameter $\beta$ is a non-negative integer, and TBS($x,y$) shows the transport block size as a function of number of assigned PRBPs, i.e. $x$, and TBS index, i.e. $y$.  Algorithm \ref{alg1} first satisfies the minimum requirements of all machine  nodes. Then, it  reimburses the extra energy consumptions for the helper nodes. Finally, it adopts the proposed algorithm\footnote{More efficient scheduling algorithms for network lifetime maximization will be appeared in \cite{jsac2}.} in \cite{vtc} to allocate the remaining PRBPs.

 \begin{algorithm}[t!]\label{alg1}
\nl $c_i^{min}(1+\beta n_i)\to c_i^{in},\quad \forall i\in \pounds_r$\;
\nl \For{$i\in \pounds_r$}{
\nl \For{$j=1:c_i^{in}$}{
- From Table 7.1.7.2.1-1 of \cite{3gpp1}, derive the lowest $\delta_i$ which satisfies $D_i\le {\text{TBS}}(j,\delta_i)$. Call it $\delta_i^j$\;
- Derive the corresponding transmit power from \eqref{pil}, i.e.  $P_i|_{\delta_i=\delta_i^j,c_i=j}\to P_i^j$\;
- Derive the expected lifetime from \eqref{lil}, i.e. $L_i|_{P_i=P_i^j}\to L_i^{j}$\;
-\If{$P_i^j>P_{max}$ }{ $0\to L_i^{j}$\;}
}
\nl $c_i=\arg\hspace{1mm}\max_{j\in \{1,\cdots,c_i^{in}\}}\hspace{1mm} L_i^j $\;
}
\nl $c_t - \sum_{i\in \pounds_r}c_i \to c_t^n$\;
\nl Use Algorithm 1 in \cite{vtc} and allocate the remaining $c_t^n$ PRBPs to the machine nodes\;
\nl \Return $c_i$\
 \caption{Cooperation-incentive scheduling for LTE networks}
\end{algorithm}

\section{Interference-aware grouped M2M communications}\label{int}
Assume a single-cell in which one\footnote{As in the uplink transmission of cellular networks each resource block is allocated only to one node at each time, the interference from intra-group communications only affects at most one node at each time. Then, here we only need to consider one primary user in the system. The same approach has been used in \cite{serveh}. } Primary
User (PU) sends data on the uplink channel to the BS with prior authentication and resource reservation. This primary user, which is randomly deployed in the cell, can be a GR, human user, or an isolated  machine-type device. In the same time, $M$ groups of machine-type devices are active and reusing the PU's uplink resources for intra-group communications. The aggregate interference at the BS from machine nodes may degrade the QoS for the PU transmission over the same uplink resource, as it is investigated in \cite{serveh} for the case of device-to-device communications.  To model the received interference at the BS we need the transmit power of machine devices and their pathloss at the base station. Let denote the transmit power of machine nodes for intra-group communications as $P_t^m$. Since the location of machine nodes is assumed to be unknown to the BS, the distance between a node to the BS as well as its corresponding pathloss is a random variable.  
 We transform the Signal to Interference and Noise Ratio (SINR) constraint for the PU, $\gamma_{pu}$, to the aggregated interference constraint as follows:
\begin{align}
\gamma_{th}&\le \gamma_{pu}\nonumber\\
 \Rightarrow  \gamma_{th}&\le \frac{P_t^{pu}G_{cb}}{\sum\nolimits_{i=1}^{M} P_t^m G_{ib}+N_0}\nonumber\\
 \Rightarrow   \sum\nolimits_{i=1}^{M} P_t^m G_{ib}&\le  \frac{P_t^{pu}G_{cb}}{\gamma_{th}}-N_0,  \label{co}
\end{align}
where $\gamma_{th}$ is the predetermined threshold SINR, $G_{cb}$ the instantaneous  gain of the PU-BS link, $P_t^{pu}$ the transmit power of the PU, and $N_0$ the noise power at the BS. Also, $G_{ib}=g_{ib}|h_{ib}|^2$ is the instantaneous channel gain between the transmitting-node in group $i$ and the base station, $g_{ib}=\frac{c}{d_{ib}^{\delta}}$ the distance-dependent pathloss, $d_{ib}$ the corresponding distance, $c$ a constant, $\delta$ the pathloss exponent, and $|h_{ib}|$ is the Rayleigh fading with ${\rm{\bf E}}_h\{ |h_{ib}|^2\}=1$ where ${\bf E}_x$ stands for expectation over $x$. We investigate the long-term average of the interference constraint in \eqref{co} as follows:
\begin{equation}\label{lh}
{\rm{{\bf {E}}}}_G \big\{\sum\nolimits_{i=1}^{M} P_t^m G_{ib}\big\}\le {\rm{{\bf {E}}}}_G \{ \frac{P_t^{pu}G_{cb}}{\gamma_{th}}-N_0 \}. 
\end{equation}
Using ${\rm{{\bf {E}}}}_G\{G_{cb}\}={c}{|d_{cb}|^{-\delta}}$, where $d_{cb}$ is the known distance between CU and the BS, we recalculate the RHS of \eqref{lh} by defining $I_{th}$ as the threshold interference, as follows:
$${\rm{{\bf {E}}}}_G \{ \frac{P_t^{pu}G_{cb}}{\gamma_{th}}-N_0 \}=\frac{P_t^{pu}c}{\gamma_{th}d_{cb}^{\delta}}-N_0 =I_{th}.$$
As channel gains are non-negative, we move the expectation inside the summation and simplify the expression as:
\begin{align}
& P_t^m\sum\limits_{i=1}^{M}  {\rm\bf E}_{G}\{G_{ib}\}\le I_{th}  \hspace{2mm}\\
 &\Rightarrow P_t^m\sum\limits_{i=1}^{M}  {\rm\bf E}_{G}\{g_{ib}|h_{ib}|^2\}\le I_{th}\nonumber\\
&\mathop  \Rightarrow \limits^{(a)}  P_t^m\sum\nolimits_{i=1}^{M}  {\rm\bf E}_{g}\{g_{ib}\}{\rm\bf E}_{h}\{|h_{ib}|^2\}\le I_{th}\nonumber\\
& \Rightarrow c P_t^m\sum\nolimits_{i=1}^{M}  {\rm\bf E}_{d}\{d_{ib}^{-\delta}\}\le I_{th}\hspace{2mm}\nonumber\\&\Rightarrow P_t^m M\le \frac{I_{th}}{c {\rm\bf E}_{d}\{d_{ib}^{-\delta}\}}\label{tr}
\end{align}
 where $(a)$ is due to the fact that pathloss and fading are independent. One sees a feasible tradeoff in $\eqref{tr}$  between number of active groups and the transmit power for intra-group communications.  To simplify the tradeoff expression in \eqref{tr}, we need to calculate the ${\rm\bf E}_{d}\{d_{ib}^{-\delta}\}$ term. We assume grouped machine-nodes are distributed uniformly in the cell with distance $ d_n\le d_{ib}\le d_x$ from the BS. Then, the probability distribution function of $d_{ib}$ is $\frac{2d_{ib}}{d_x^2}$ and we have:
 \begin{align}
 {\rm\bf E}_{d}\{d_{ib}^{-\delta}\}=&\int\nolimits_{{d_n}}^{{d_x}}\frac{1}{d_{ib}^{\delta}}\frac{2d_{ib}}{d_{x}^2} \text{ d} d_{ib}
  =\left\{ \begin{array}{l}
\frac{2}{d_x^2}\ln \frac{d_x}{d_n} \qquad \hspace{7mm} \delta= 2\\
\\
\frac{2(d_x^{2-\delta}-d_n^{2-\delta})}{d_x^2(2-\delta)} \hspace{2mm}\qquad  \text{O.W.}
\end{array} \right.\nonumber
 \end{align}
 Finally, we can rewrite the tradeoff expression in \eqref{tr}, for $\delta >2$, as follows:
\begin{align}
 P_t^m\times  M &\le \frac{I_{th}d_x^2(2-\delta)}{2c(d_x^{2-\delta}-d_n^{2-\delta})} = \frac{d_x^2(\delta-2)}{\frac{1}{d_n^{\delta-2}}-\frac{1}{d_x^{\delta-2}}} \times \frac{I_{th}}{2c}\label{tra}
\end{align} 
It is evident in \eqref{tra} that given the transmit power for intra-group communications $P_t^m$, by increase in the pathloss-exponent or $d_n$, the maximum allowable number of concurrent active groups in the cell increases.
Assume semi circular-shaped clusters are formed in the cell with diameter $\Delta$ such that the average received power from the group-edge machine device is $P_r^m$ at the GR which is located $\Delta$ meters far from the transmitter in the worse case. Then the transmit power of machine devices will be at most $P_t^m={P_r^m}\Delta^{\delta_m}/{c_m}$, where $c_m$ is a constant and $\delta_m$ is the pathloss exponent for the intra-group communications. Now one can rewrite the expression in \eqref{tra} as follows:
\begin{align}
 \Delta^{\delta_m}\times  M &\le \frac{d_x^2(\delta-2)}{\frac{1}{d_n^{\delta-2}}-\frac{1}{d_x^{\delta-2}}} \times \frac{c_m I_{th}}{2c_h P_r^m}.\label{trad}
\end{align} 
  The inequality in \eqref{trad} presents an interesting  tradeoff between the maximum number and  radius of groups that can be supported in a given cell when there is a QoS requirement for the primary user. Specially, one can see that by doubling the radius of the groups, the number of active groups must decrease 4 times when $\delta_m=2$.


 \begin{table}[t!]
\centering \caption{Simulation Parameters}\label{par}
\begin{tabular}{p{4 cm}p{4 cm}}\\
\toprule[0.5mm]
Parameter&Value\\
\midrule[0.5mm]
Cell inner and outer radius & 50 and 450 m \\
Number of nodes, $\mathcal L$&40 \\
Pathloss model & adopted from \cite{6240186}\\
Total resources, $c_t \tau_r$&$2.5\mathcal L \times\max_{i\in \pounds}\{\tau_i^{min}\} $\\
PSD of noise, $N_0$& -121 dBW\\ 
Threshold SINR for PU, $\gamma_{th}$&0, 2 dB\\ 
Static Energy cons., $E_s^i$& 50 $\mu$Joule\\
Circuit power, $P_c$& 1 mW\\
$P_t^{cu}=P_{\max}$& 1 W\\
Full battery capacity& 250 Joule\\
\bottomrule[0.5mm]
\end{tabular}
\end{table}

\section{Performance Evaluation}\label{sim}
In this section, we evaluate the system performance. The testbed for our simulations is implemented in Matlab and consists of   a single-cell, one BS at the cell-center, and $\mathcal L$ battery-driven machine devices which are randomly deployed according to a spatial Poisson point process (SPPP) in the cell with a minimum distance of 50 m from the base station. When cooperation among machine nodes is enabled, the GRs schedule their group members to send their data before the allocated resource pool starts by reusing the uplink cellular resources (inband underlay D2D). As the interference from PUs can significantly degrade the reliability of intra-group communications, the machine nodes 
 sense the carrier before data transmission. The other simulation parameters can be found in Table \ref{par}. 

\begin{figure}[!t]
\centering
\includegraphics[trim={1cm 0.1cm .5cm .2cm},clip,width=5in]{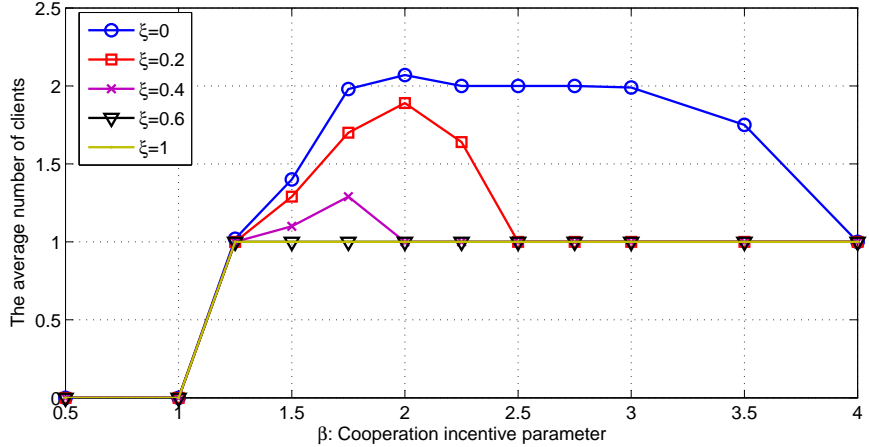}
\caption{ Degrees of motivation for cooperation versus $\beta$. } \label{dom}
\end{figure}

Fig. \ref{dom} compares the degrees of motivation for cooperation among machine nodes for different $\beta$ and $\xi$ values   where $\beta$ is the cooperation incentive parameter and $\xi$ is the average $E_h$ to $E_s^i$ ratio. Given the $\beta$ value, each machine node determines its optimal number of clients by solving the optimization problem in \eqref{op18}. From Fig. \ref{dom} one sees as the cost per client increases, the motivation of machine nodes for cooperation decreases. Also, we see that an increase in the allocated time for data transmission, which is achieved by increasing $\beta$, doesn't necessary lead to a higher level of motivation for cooperation. This is due to the fact that machine nodes have also circuit energy consumptions.  In most previous works on energy efficient MTC scheduling, only the transmit energy is considered and the other energy consumptions by the operation of electronic circuits, which are comparable or more dominant than  the energy consumption for reliable data transmission \cite{guo,joz}, are neglected \cite{poshti,ra,coex}. Here, one sees that such an assumption may result in misleading grouping guidelines.

\begin{figure}[!t]
\centering
\includegraphics[trim={1cm .1cm .5cm .2cm},clip,width=5in]{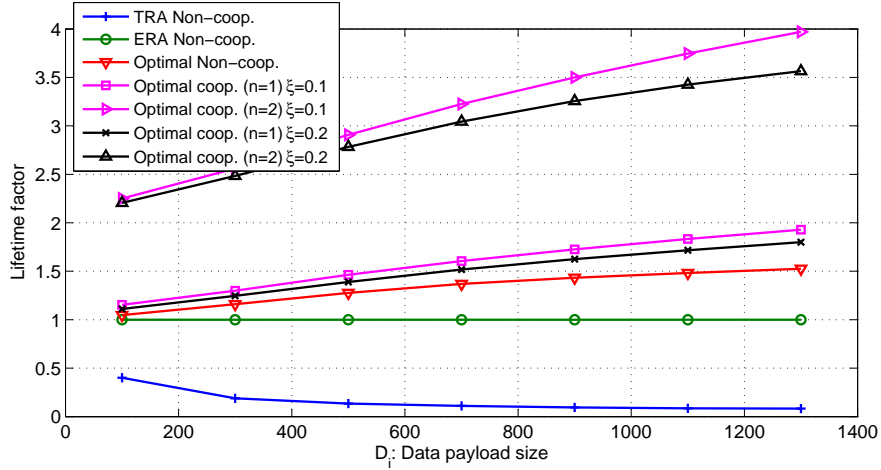}
\caption{ FED network lifetime  for different scheduling schemes} \label{fed}
\end{figure}

The network lifetime performance of the proposed joint scheduling and power control scheme and its augmented version which utilizes cooperation among nodes are investigated in Fig. \ref{fed}. 
As a benchmark, we compare lifetime performance of the proposed schemes with the results of the following schemes: (i) Equal Resource Allocation (ERA) in which resource elements are distributed equally among the nodes; and (ii) Throughput-aware Resource Allocation (TRA) in which, the closer nodes to the BS have priority in resource allocation. In our simulation model, we perform the scheduling at the reference time $t_0$ when the remaining energy of each machine device is a random value between zero and full battery capacity.
Also, the lifetime results in this figure are normalized to the lifetime performance of the ERA scheme and are shown as {\it lifetime factor} in order to provide a comparative analysis rather than a quantitative analysis. From Fig. \ref{fed}, it is evident that the FED network lifetime is  extended using the optimal non-cooperative solution in \eqref{tis}. Also, one sees that significant improvement in network lifetime can be achieved by triggering cooperation among nodes. For example, when machine nodes construct groups of size 3, i.e. $n\buildrel \Delta \over = n_i=2, \hspace{1mm}\forall i\in \pounds_r$, the resulted network lifetime from our proposed scheme in \eqref{ti*} is approximately 200$\%$ more than the case of ERA scheme.  One can also see that as the energy consumption in receiving data from group members increases, i.e. $\xi$ increases, the lifetime benefit from grouping decreases. Then, lifetime-aware data transmission schemes for intra-group communications, e.g. the proposed schemes in \cite{gsip,jsac}, are important for network lifetime maximization. Furthermore, one sees that the lifetime factor increases in the data payload size. This is due to the fact that the transmit power, and hence the energy consumption, increase in data payload size.

\subsection{Performance evaluation in the context of LTE }\label{simL}
\begin{figure}[!t]
\centering
\includegraphics[trim={1cm .1cm 0cm .2cm},clip,width=5in]{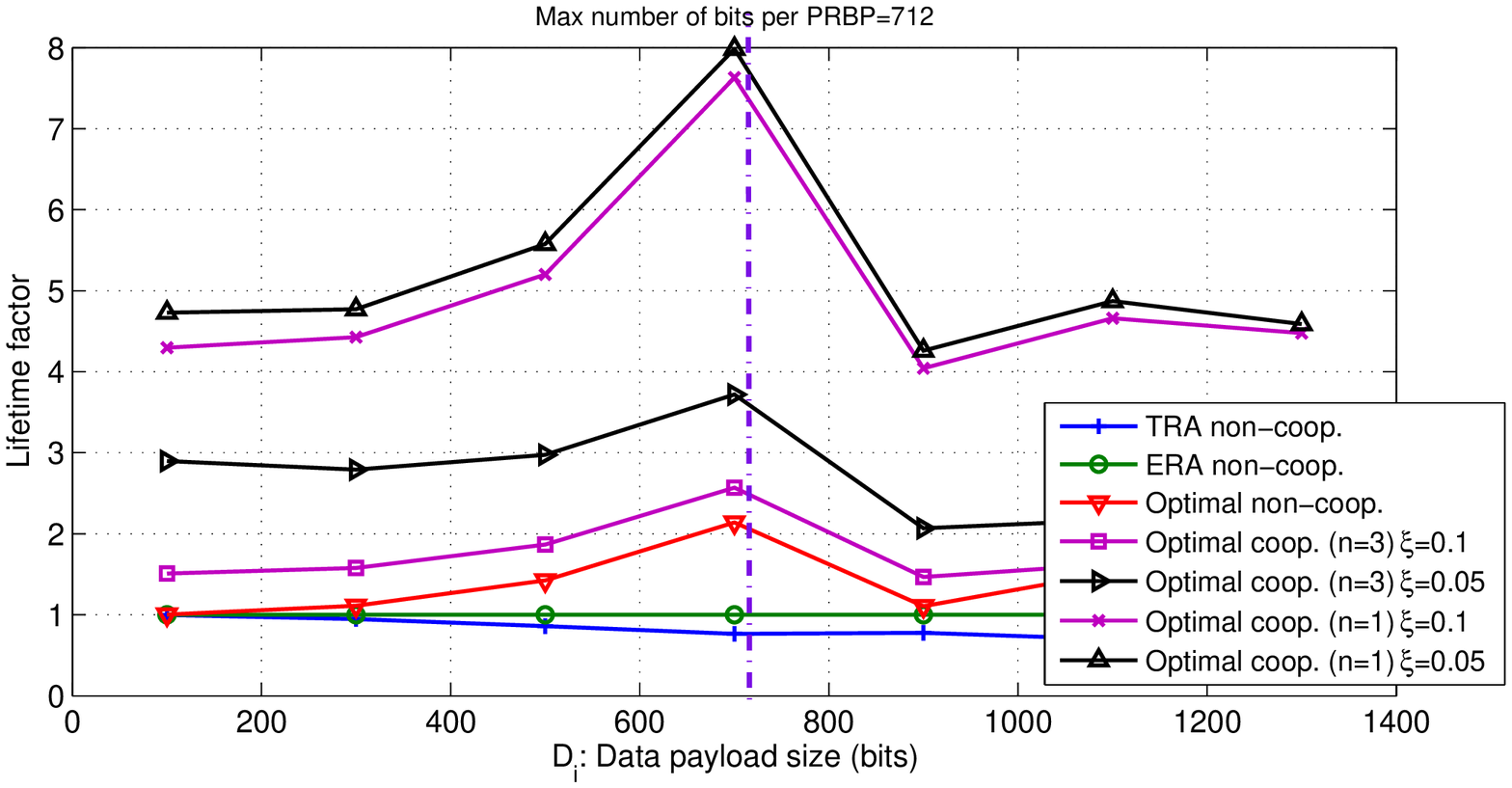}
\caption{ FED network Lifetime  for different scheduling schemes} \label{fedL}
\end{figure}

Here, we implement the uplink transmission of a single cell, multi-user 3GPP LTE network in Matlab.  Machine nodes are randomly deployed in the cell with a minimum distance of 50 m from the base station. The other simulation parameters can be found in Table \ref{par} unless specified in Table \ref{parL}. Our aim here is to investigate how the constraints on the transport block size, discrete modulation and coding indexes, and discrete time/frequency resources in the existing 3GPP LTE networks  can influence our proposed scheduling solutions. 
 The FED  lifetime performance of the network is presented in Fig. \ref{fedL}. In comparison with Fig. \ref{fed}, ones sees that  the lifetime curves are not smooth any more. This bell-shaped behavior can be argued as follows. As one sees, the lifetime curves are increasing in $D_i$ until $D_i=712$, which is the maximum TBS per one PRBP. When $D_i$ is relatively small, e.g. 100 bits, the required TBS index from Fig. \ref{tbs} is found to be 7, which results in a relatively low transmit power. When $D_i$ is sufficiently large, e.g. 700 bits, the required TBS index from Fig. \ref{tbs}  is found to be 26, which results in a relatively high transmit power. As the energy consumption, and hence the lifetime degradation, in large $D_i$ values are more critical, one sees that the lifetime factor increases in $D_i$ in our proposed solutions, and decreases in $D_i$ for the throughput-aware solution which is not designed to be lifetime-aware. When $D_i$ exceeds 712, the $c_i^{min}=\left\lceil {\frac{{{D_i}}}{{712}}} \right\rceil$ increases from 1 to 2. Now, if $D_i=712+d$ where $\frac{d}{712}$ is relatively small, the situation becomes similar to the case where $D_i$ is small, and hence, the difference between the proposed schemes and the ERA is not significant. Then, one sees a significant decrease in the lifetime-factor after $D_i=712$. Finally, one sees that when $n=3$, the lifetime factor is lower than the case when $n=1$. This is in accordance with our results in Fig. \ref{dom}, where we have found that increasing the group size may decrease the lifetime-factor as the BS can not reimburse all energy consumptions by the helper nodes.

\begin{figure}[!t]
\centering
\includegraphics[trim={1cm .05cm .5cm .2cm},clip,width=5in]{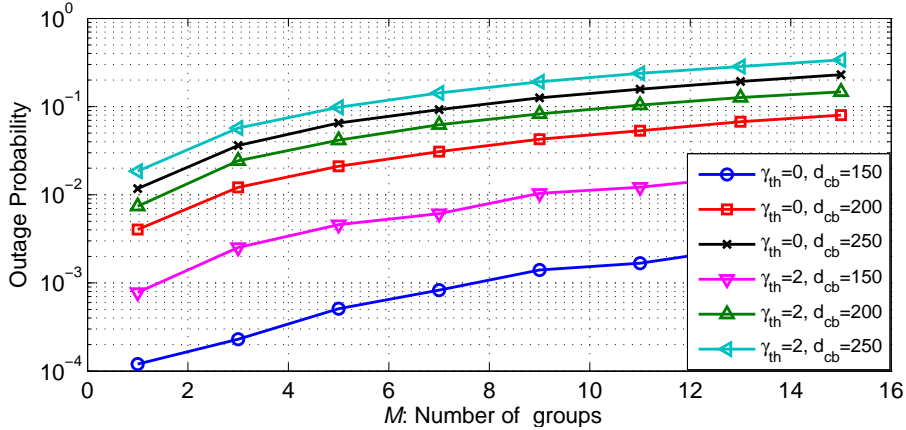}
\caption{ QoS for primary user versus different number of groups in the cell.} \label{outage}
\end{figure}

Fig. \ref{outage} depicts the outage probability in uplink transmission of a primary user versus different numbers of active groups in the cell, $M$. Given $M$, the maximum transmit power of machine nodes for intra-group communications is controlled by  \eqref{tra}. From Fig. \ref{outage}, one sees that when PU is far away from the BS, e.g. $d_{cb}=250 m$, even a small number of concurrent reuses of the uplink resource degrades the outage performance significantly. When PU is located closer to the BS, e.g. $d_{ib}=150 m$, one sees that 9 concurrent reuses of the uplink resource degrades the outage performance at most one order of magnitude. These results show that the  coexistence of machine- and human-oriented traffic should be considered in next generations of cellular networks to guarantee the QoS for primary users while enabling group-based M2M communications.

\begin{table}[t!]
\centering \caption{Simulation Parameters for an LTE Network  \cite{3gpp1}}\label{parL}
\begin{tabular}{p{8 cm}p{3.9 cm}}\\
\toprule[0.5mm]
Parameter&Value\\
\midrule[0.5mm]
Data payload size, $D_i$& [100,1300]\\
Downlink pathloss compensation factor, $\beta_i$&0.91\\
Total number of PRBPs, $c_t$&60 $\times c^{min}$  \\
Maximum transmit power, $P_{max}$& -6 dB\\
 Transmission time interval, TTI& 1 msec\\
$\text{SNR}_{\text{target}}$&-5 dB\\
$N_s $& 12 \\
Transport block size& Table 7.1.7.2.1-1 of \cite{3gpp1}\\
\bottomrule[0.5mm]
\end{tabular}
\end{table}
%

\subsection{How does each network entity benefit from cooperation? }
In this section we analyze how each entity of the network, i.e. Group Member (GM), GR, and BS, benefits from cooperation, when the proposed rewarding mechanism is utilized in the network. The potential energy-saving in data transmission for  group members happens because of relatively shorter distance between a GM and its respective GR than the distance between a GM and the BS. Also, local synchronization and coordination with GR for data transmission wastes less energy than the case of computationally complex synchronization with the BS \cite{aza}, especially when the BS is heavily loaded or it must serve a massive number of nodes. The potential energy-saving for GRs comes from the proposed incentive algorithm in this work where each GR has this opportunity to transmit $(n_i+1) {\hat D}$ bits of data in $(\beta n_i+1)c_i^{min}$ resource elements instead of transmitting $D_i$ bits of data in $c_i^{min}$ resource elements. Furthermore, the GRs won't suffer from potential congestion in the RACH connection due to less number of nodes which need direct connection to the BS.   The potential benefit for the BS comes from dealing with lower number of nodes, in contradictory with the case of direct access, where the BS must sustain a massive number of short-lived sessions, while it has been designed and optimized for a small number of long-lived sessions. Last but not least, being energy-efficient in supporting machine-type communications also let the network to become green because with a lower number of nodes, each BS can handover its potential arriving  users to the neighbor cells and go to the sleep mode which results in a huge energy saving for the access network \cite{sleep}.

\section{Conclusion}
In this paper, a lifetime- and interference-aware incentive mechanism for triggering cooperation among machine-type devices underlying cellular networks is investigated. Aiming at maximizing  network lifetime and offloading the BS, we present a cooperation incentive scheme which  reimburses the extra energy consumptions for the helper nodes. We then analyze the impact of underlying intra-group communications on the uplink transmission of primary users. Our theoretical and simulation results show that  significant improvement in the network lifetime can be achieved by  triggering cooperation among nodes, while guaranteeing   QoS for the primary users. 


\ifCLASSOPTIONcaptionsoff
  \newpage
\fi

\bibliographystyle{IEEEtran}
\bibliography{bibl}

\begin{thebibliography}{10}
\providecommand{\url}[1]{#1}
\csname url@samestyle\endcsname
\providecommand{\newblock}{\relax}
\providecommand{\bibinfo}[2]{#2}
\providecommand{\BIBentrySTDinterwordspacing}{\spaceskip=0pt\relax}
\providecommand{\BIBentryALTinterwordstretchfactor}{4}
\providecommand{\BIBentryALTinterwordspacing}{\spaceskip=\fontdimen2\font plus
\BIBentryALTinterwordstretchfactor\fontdimen3\font minus
  \fontdimen4\font\relax}
\providecommand{\BIBforeignlanguage}[2]{{%
\expandafter\ifx\csname l@#1\endcsname\relax
\typeout{** WARNING: IEEEtran.bst: No hyphenation pattern has been}%
\typeout{** loaded for the language `#1'. Using the pattern for}%
\typeout{** the default language instead.}%
\else
\language=\csname l@#1\endcsname
\fi
#2}}
\providecommand{\BIBdecl}{\relax}
\BIBdecl

\bibitem{fair}
I.~H. Jafri, M.~S. Soomro, and R.~A. Khan, ``{ICT} in distance education:
  Improving literacy in the province of sindeh pakistan,'' \emph{The Sindh
  University Journal of Education}, vol.~40, 2011.

\bibitem{green}
F.~Mattern, T.~Staake, and M.~Weiss, ``{ICT} for green: how computers can help
  us to conserve energy,'' in \emph{Proceedings of the 1st international
  conference on energy-efficient computing and networking}.\hskip 1em plus
  0.5em minus 0.4em\relax ACM, 2010, pp. 1--10.

\bibitem{hel}
D.~Miorandi, S.~Sicari, F.~D. Pellegrini, and I.~Chlamtac, ``Internet of
  things: Vision, applications and research challenges,'' \emph{Ad Hoc
  Networks}, vol.~10, no.~7, pp. 1497 -- 1516, 2012.

\bibitem{rico}
\BIBentryALTinterwordspacing
Ericsson, ``More than 50 billion connected devices,'' Tech. Rep., 2011.
  [Online]. Available: \url{http://www.ericsson.com}
\BIBentrySTDinterwordspacing

\bibitem{m2m1}
K.~C. Chen and S.~Y. Lien, ``Machine-to-machine communications: Technologies
  and challenges,'' \emph{Ad Hoc Networks}, vol.~18, pp. 3 -- 23, 2014.

\bibitem{itu}
L.~Srivastava, T.~Kelly, and {{et al.}}, ``The internet of things,''
  International Telecommunication Union, Tech. Rep.~7, Nov 2005.

\bibitem{w_sony}
Ericsson, Huawei, NSN, and et~al., ``A choice of future {M2M} access
  technologies for mobile network operators,'' Tech. Rep., March 2014.

\bibitem{w_sam}
{Samsung Electronics}, ``{5G} vision,'' Tech. Rep., 2015.

\bibitem{ghav}
F.~Ghavimi and H.~H. Chen, ``{M2M} communications in {3GPP LTE/LTE-A} networks:
  Architectures, service requirements, challenges, and applications,''
  \emph{IEEE communications Surveys and Toturials}, vol.~17, no.~2, pp.
  525--549, 2015.

\bibitem{cisco}
{Samsung Electronics}, ``Cisco visual networking index: Global mobile data
  traffic forecast update, 2014–2019,'' Tech. Rep., March 2015.

\bibitem{llte}
{3GPP TR 37.868 V11.0.0}, ``Study on {RAN} improvements for machine-{T}ype
  communications,'' Tech. Rep., Sep 2011.

\bibitem{8l}
M.~Beale, ``Future challenges in efficiently supporting {M2M} in the {LTE}
  standards,'' in \emph{IEEE Wireless Communications and Networking Conference
  Workshops}, April 2012, pp. 186--190.

\bibitem{9l}
K.~Zheng, F.~Hu, W.~Wang, W.~Xiang, and M.~Dohler, ``Radio resource allocation
  in {LTE}-advanced cellular networks with {M2M} communications,'' \emph{IEEE
  communications Magazine}, vol.~50, no.~7, pp. 184--192, July 2012.

\bibitem{how}
R.~Ratasuk, A.~Prasad, Z.~Li, A.~Ghosh, and M.~Uusitalo, ``Recent advancements
  in {M2M} communications in {4G} networks and evolution towards {5G},'' in
  \emph{IEEE International Conference on Intelligence in Next Generation
  Networks}, 2015, pp. 52--57.

\bibitem{m2monlte}
M.~Gerasimenko, V.~Petrov, O.~Galinina, S.~Andreev, and Y.~Koucheryavy,
  ``Impact of machine-type communications on energy and delay performance of
  random access channel in {LTE}-advanced,'' \emph{Transactions on Emerging
  Telecommunications Technologies}, vol.~24, no.~4, pp. 366--377, 2013.

\bibitem{how1}
M.~Islam, A.~Taha, and S.~Akl, ``A survey of access management techniques in
  machine-type communications,'' \emph{IEEE communications Magazine}, vol.~52,
  no.~4, pp. 74--81, 2014.

\bibitem{sched}
A.~Gotsis, A.~Lioumpas, and A.~Alexiou, ``{M2M} scheduling over {LTE}:
  Challenges and new perspectives,'' \emph{IEEE Vehicular Technology Magazine},
  vol.~7, no.~3, pp. 34--39, Sept. 2012.

\bibitem{andres}
{ETSI TS 102 690 V1.1.1}, ``Machine-to-machine communications ({M2M});
  functional architecture,'' International Telecommunication Union, Tech. Rep.,
  October 2011.

\bibitem{acb1}
S.~Duan, V.~Shah-Mansouri, and V.~W. Wong, ``Dynamic access class barring for
  {M2M} communications in {LTE} networks,'' in \emph{IEEE Global Communications
  Conference}, 2013, pp. 4747--4752.

\bibitem{acb2}
M.~Hasan, E.~Hossain, and D.~Niyato, ``Random access for machine-to-machine
  communication in {LTE}-advanced networks: issues and approaches,'' \emph{IEEE
  communications Magazine}, vol.~51, no.~6, pp. 86--93, 2013.

\bibitem{pri}
J.~P. Cheng, C.~h. Lee, and T.~M. Lin, ``Prioritized random access with dynamic
  access barring for {RAN} overload in {3GPP LTE-A} networks,'' in \emph{IEEE
  GLOBECOM Workshops}, 2011, pp. 368--372.

\bibitem{pri1}
T.~M. Lin, C.~H. Lee, J.~P. Cheng, and W.~T. Chen, ``{PRADA}: prioritized
  random access with dynamic access barring for {MTC} in {3GPP LTE-A}
  networks,'' \emph{IEEE Transactions on Vehicular Technology}, vol.~63, no.~5,
  pp. 2467--2472, 2014.

\bibitem{con1}
Y.~Chen and W.~Wang, ``Machine-to-machine communication in {LTE-A},'' in
  \emph{IEEE 72nd Vehicular Technology Conference}, 2010, pp. 1--4.

\bibitem{con2}
R.~P. Jover and I.~Murynets, ``Connection-less communication of {IoT} devices
  over {LTE} mobile networks,'' in \emph{12th Annual IEEE International
  Conference on Sensing, Communication, and Networking}, 2015, pp. 247--255.

\bibitem{r10}
A.~R. Prasad, ``{3GPP SAE/LTE} security,'' \emph{NIKSUN WWSMC}, 2011.

\bibitem{r11}
3rd Generation Partnership Project; Technical Specification Group~Services and
  S.~Aspects, ``Study on core network overload and solutions,'' \emph{3GPP TR
  23.843,” vol. v0.7.0}, 2012.

\bibitem{sca}
J.~Jermyn, R.~P. Jover, I.~Murynets, M.~Istomin, and S.~Stolfo, ``Scalability
  of machine to machine systems and the internet of things on {LTE} mobile
  networks,'' Tech. Rep.

\bibitem{dtx}
J.~Wu, T.~Zhang, Z.~Zeng, and H.~Chen, ``Study on discontinuous reception
  modeling for {M2M} traffic in {LTE-A} networks,'' in \emph{IEEE International
  Conference on Communication Technology}, Nov 2013, pp. 584--588.

\bibitem{sach}
T.~Tirronen, A.~Larmo, J.~Sachs, B.~Lindoff, and N.~Wiberg, ``Reducing energy
  consumption of {LTE} devices for machine-to-machine communication,'' in
  \emph{IEEE Globecom Workshops}, Dec 2012, pp. 1650--1656.

\bibitem{dtx1}
D.~P. Van, B.~P. Rimal, and M.~Maier, ``Power-saving scheme for {PON LTE-A}
  converged networks supporting {M2M} communications,'' in \emph{IEEE
  International Conference on Ubiquitous Wireless Broadband}, 2015, pp. 1--5.

\bibitem{gsip}
A.~Azari and G.~Miao, ``Energy efficient {MAC} for cellular-based {M2M}
  communications,'' in \emph{2nd IEEE Global Conference on Signal and
  Information Processing}, 2014.

\bibitem{sachs}
J.~Sachs, N.~Beijar, P.~Elmdahl, J.~Melen, F.~Militano, and P.~Salmela,
  ``Capillary networks--a smart way to get things connected,'' \emph{Ericsson
  Review, September}, vol.~9, 2014.

\bibitem{gw1}
V.~Mišić, J.~Mišić, X.~Lin, and D.~Nerandzic, in \emph{Ad-hoc, Mobile, and
  Wireless Networks}.\hskip 1em plus 0.5em minus 0.4em\relax Springer Berlin
  Heidelberg, 2012, vol. 7363, pp. 413--423.

\bibitem{grlaya}
A.~Laya, L.~Alonso, J.~Alonso-Zarate, and M.~Dohler, ``Green {MTC, M2M},
  internet of things,'' \emph{Green Communications: Principles, Concepts and
  Practice}, pp. 217--236, 2015.

\bibitem{d2ddo}
N.~Pratas and P.~Popovski, ``Zero-outage cellular downlink with fixed-rate
  {D2D} underlay,'' \emph{IEEE Transactions on Wireless Communications},
  vol.~14, no.~7, pp. 3533--3543, July 2015.

\bibitem{d2dagg}
G.~Rigazzi, N.~Pratas, P.~Popovski, and R.~Fantacci, ``Aggregation and trunking
  of {M2M} traffic via {D2D} connections,'' in \emph{IEEE International
  Conference on Communications}, June 2015, pp. 2973--2978.

\bibitem{d2dmh}
G.~Rigazzi, F.~Chiti, R.~Fantacci, and C.~Carlini, ``Multi-hop {D2D} networking
  and resource management scheme for {M2M} communications over {LTE-A}
  systems,'' in \emph{International Wireless Communications and Mobile
  Computing Conference}, 2014, pp. 973--978.

\bibitem{vtc}
A.~Azari and G.~Miao, ``Lifetime-aware scheduling and power control for {M2M}
  communications over {LTE} networks,'' in \emph{IEEE WCNC}, 2015.

\bibitem{jsac}
A.~Azari, G.~Miao, and T.~Hwang, ``${E}^2$-{MAC}: Energy efficient medium
  access for massive {M2M} communications,'' \emph{IEEE Transactions on
  Communications}, submitted.

\bibitem{coex}
A.~Aijaz, M.~Tshangini, M.~Nakhai, X.~Chu, and A.-H. Aghvami,
  ``Energy-efficient uplink resource allocation in {LTE} networks with
  {M2M/H2H} co-existence under statistical {QoS} guarantees,'' \emph{IEEE
  Transactions communications}, vol.~62, no.~7, pp. 2353--2365, July 2014.

\bibitem{poshti}
H.~S. Dhillon, H.~C. Huang, H.~Viswanathan, and R.~A. Valenzuela,
  ``Power-efficient system design for cellular-based machine-to-machine
  communications,'' \emph{IEEE Transactions Wireless communications}, vol.~12,
  no.~11, pp. 5740--5753, November 2013.

\bibitem{ra}
H.~Safdar, N.~Fisal, R.~Ullah, W.~Maqbool, and Z.~Khalid, ``Resource allocation
  for uplink {M2M} communication in multi-tier network,'' in \emph{IEEE
  Malaysia International Conference on Communications}, 2013, pp. 538--543.

\bibitem{sysreq}
\BIBentryALTinterwordspacing
{3GPP TS 22.368 V13.1.0}, ``Service requirements for machine-type
  communications,'' Tech. Rep., 2014. [Online]. Available:
  \url{http://www.3gpp.org/ftp/Specs/archive/22\_series/22.368/22368-d10.zip}
\BIBentrySTDinterwordspacing

\bibitem{m2meee}
H.~Cho, ``Machine to machine {(M2M)} communications technical report,'' 2011,
  {IEEE} 802.16 Broadband Wireless Access Working Group.

\bibitem{mas}
S.~Y. Lien and K.~C. Chen, ``Massive access management for {QoS} guarantees in
  {3GPP} machine-to-machine communications,'' \emph{IEEE communications
  Letters}, vol.~15, no.~3, pp. 311--313, March 2011.

\bibitem{ana}
A.~G. Gotsis, A.~S. Lioumpas, and A.~Alexiou, ``Analytical modeling and
  performance evaluation of realistic time-controlled {M2M} scheduling over
  {LTE} cellular networks,'' \emph{Transaction Emerging Telecommunications
  Technologies}, vol.~24, no.~4, pp. 378--388, 2013.

\bibitem{grp1}
------, ``Evolution of packet scheduling for machine-type communications over
  {LTE}: Algorithmic design and performance analysis,'' in \emph{2012 IEEE
  Globecom Workshops}, 2012, pp. 1620--1625.

\bibitem{grp3}
M.~Giluka and et~al., ``Class based priority scheduling to support machine to
  machine communications in {LTE} systems,'' in \emph{20th National Conference
  on Communications}, 2014, pp. 1--6.

\bibitem{clasque}
T.~Kwon and J.-W. Choi, ``Multi-group random access resource allocation for
  {M2M} devices in multicell systems,'' \emph{IEEE communications Letters},
  vol.~16, no.~6, pp. 834--837, 2012.

\bibitem{tree}
Y.~Zhang, ``Tree-based resource allocation for periodic cellular {M2M}
  communications,'' \emph{IEEE Wireless communications Letters}, vol.~3, no.~6,
  pp. 621--624, Dec 2014.

\bibitem{reli}
G.~Madueño, C.~Stefanovic, and P.~Popovski, ``Reliable reporting for massive
  {M2M} communications with periodic resource pooling,'' \emph{IEEE Wireless
  communications Letters}, vol.~3, no.~4, pp. 429--432, Aug 2014.

\bibitem{wcnc}
A.~Azari and G.~Miao, ``Lifetime-aware scheduling and power control for
  cellular-based {M2M} communications,'' in \emph{IEEE WCNC}, 2015.

\bibitem{m2mm2}
T.~Taleb and A.~Ksentini, ``On alleviating {MTC} overload in {EPS},'' \emph{Ad
  Hoc Networks}, vol.~18, pp. 24 -- 39, 2014.

\bibitem{mc1}
C.~Y. Ho and C.~Y. Huang, ``Energy-saving massive access control and resource
  allocation schemes for {M2M} communications in {OFDMA} cellular networks,''
  \emph{IEEE Wireless communications Letters}, vol.~1, no.~3, pp. 209--212,
  June 2012.

\bibitem{istiak}
M.~I. Hossain, A.~Laya, F.~Militano, S.~Iraji, and J.~Markendahl, ``Reducing
  signaling overload: Flexible capillary admission control for dense {MTC} over
  {LTE} networks,'' in \emph{IEEE 26th Annual International Symposium on
  Personal, Indoor, and Mobile Radio Communications}, 2015, pp. 1305--1310.

\bibitem{boyd_con}
S.~Boyd and L.~Vandenberghe, \emph{Convex optimization}.\hskip 1em plus 0.5em
  minus 0.4em\relax Cambridge university press, 2004.

\bibitem{lam}
R.~Corless, G.~Gonnet, D.~Hare, D.~Jeffrey, and D.~Knuth, ``On the {LambertW}
  function,'' vol.~5, no.~1, pp. 329--359, 1996.

\bibitem{round}
P.~Raghavan and C.~D. Tompson, ``Randomized rounding: a technique for provably
  good algorithms and algorithmic proofs,'' \emph{Combinatorica}, vol.~7,
  no.~4, pp. 365--374, 1987.

\bibitem{lear}
M.~Bennis and D.~Niyato, ``A {Q}-learning based approach to interference
  avoidance in self-organized femtocell networks,'' in \emph{IEEE GLOBECOM
  Workshops}, Dec 2010, pp. 706--710.

\bibitem{3gpp}
{3GPP TS 36.300}, ``Evolved universal terrestrial radio access ({E-UTRA}) and
  evolved universal terrestrial radio access network ({E-UTRAN}); overall
  description.'' Tech. Rep., (Release 10).

\bibitem{3gpp1}
{3GPP TS 36.213}, ``Evolved universal terrestrial radio access ({EUTRA});
  physical layer procedures.'' Tech. Rep., (Release 10).

\bibitem{jsac2}
``Network lifetime maximization for cellular based m2m networks, author={Azari,
  Amin and Miao}, journal={IEEE Journal on Selected areas on Communications},
  note={submitted}.''

\bibitem{serveh}
S.~Shalmashi, G.~Miao, Z.~Han, and B.~Slimane, ``Interference constrained
  device-to-device communications,'' in \emph{{IEEE} {ICC}}, 2014.

\bibitem{6240186}
R.~Ratasuk, J.~Tan, and A.~Ghosh, ``Coverage and capacity analysis for machine
  type communications in {LTE},'' in \emph{IEEE Vehicular Technology Conference
  (spring)}, May 2012, pp. 1--5.

\bibitem{guo}
G.~Miao, N.~Himayat, G.~Li, and S.~Talwar, ``Low-complexity energy-efficient
  scheduling for {Uplink OFDMA},'' \emph{IEEE Transactions communications},
  vol.~60, no.~1, pp. 112--120, Jan. 2012.

\bibitem{joz}
T.~Tirronen, A.~Larmo, J.~Sachs, B.~Lindoff, and N.~Wiberg, ``Reducing energy
  consumption of {LTE} devices for machine-to-machine communication,'' in
  \emph{IEEE Globecom Workshops}, Dec 2012, pp. 1650--1656.

\bibitem{aza}
A.~Azari, S.~N. Esfahani, and M.~G. Roozbahani, ``A new method for timing
  synchronization in {OFDM} systems based on polyphase sequences,'' in
  \emph{IEEE 71st}.

\bibitem{sleep}
E.~Oh, K.~Son, and B.~Krishnamachari, ``Dynamic base station switching-on/off
  strategies for green cellular networks,'' \emph{IEEE Transactions on Wireless
  Communications}, vol.~12, no.~5, pp. 2126--2136, 2013.

\end{thebibliography}

\end{document}